%% file: TwoLoopKonishi.tex
\documentclass{article}
\usepackage{jheppub}
\usepackage{amsmath,amsfonts,amsthm,graphicx}
\usepackage{ulem}
\usepackage{mathrsfs}
\usepackage[abs]{overpic}
\usepackage{rotating}
\numberwithin{equation}{section}

\newcommand{\bal}{\begin{align}}
\newcommand{\eal}{\end{align}}
\newcommand{\beq}{\begin{equation}}
\newcommand{\eeq}{\end{equation}}
\newcommand\beqa{\begin{eqnarray}}
\newcommand\eeqa{\end{eqnarray}}
\newcommand\bea{\begin{array}}
\newcommand\eea{\end{array}}
\newcommand\comment[1]{{}}

\newcommand{\eq}[1]{(\ref{#1})}

\newcommand{\su}{\frak{su}}
\renewcommand{\sl}{\frak{sl}}

\renewcommand{\)}{\right)}
\renewcommand{\(}{\left(}
\renewcommand{\]}{\right]}
\renewcommand{\[}{\left[}

    \newcommand{\nn}{\nonumber}
    
    \newcommand{\COMMENT}[1]{}
    
    \newcommand{\neqa}{\nonumber\end{eqnarray}}
    \newcommand{\la}[1]{\label{#1}}

\def\[{\left[}
\def\]{\right]}

\def\<{\langle}
\def\>{\rangle}

\def\i2{\frac{i}{2}}

\def\<{\langle}
\def\>{\rangle}

\def\i2{\frac{i}{2}}

\def\tr{{\rm tr}}

\def\1h{\hat 1}
\def\2h{{\hat 2}}
\def\3h{{\hat 3}}
\def\4h{{\hat 4}}

\def\be{\begin{eqnarray}}
\def\ee{\end{eqnarray}}

    \def\<{\left\langle\,}
    \def\>{\, \right\rangle}
    \def\[{\left[}
    \def\]{\right]}

   \def\su{{\mathfrak{su}}}
   \def\sl{{\mathfrak{sl}}}

\usepackage{amssymb}
\usepackage{subfig}
\usepackage{overpic}

\title{Deeper Look into Short Strings}

\author[a,b]{Nikolay Gromov}
\author[a]{Saulius Valatka}

\affiliation[a]{Mathematics Department, King's College London,
The Strand, London WC2R 2LS, UK.}
\affiliation[b]{St.Petersburg INP, Gatchina, 188 300, St.Petersburg,
  Russia.}

\emailAdd{nikolay.gromov$\bullet$kcl.ac.uk}
\emailAdd{saulius.valatka$\bullet$kcl.ac.uk}

\abstract{
Using a recent conjecture of Basso we compute three leading nontrivial coefficients in the strong coupling
expansion of the anomalous dimensions of short operators in the $\sl_2$ sector of AdS/CFT.
We show that our results are consistent with the numerical results
obtained using the Y-system and TBA approach earlier
thus providing further support to the Y-system conjecture.
}


\usepackage{pstricks}

\usepackage{varioref}

\begin{document}

\maketitle

\section{Introduction}
For the past decade the AdS/CFT correspondence \cite{Maldacena:1997re,Gubser:1998bc,Witten:1998qj}
has been an incredible source of inspiration for theoretical physics.
Together with the integrability of the world-sheet sigma model it allows one
to obtain highly nontrivial results in four dimensional gauge theories.
A particular example is the planar ${\cal N}=4$ Super Yang-Mills theory where integrability
methods allow one to compute the spectrum of anomalous dimensions as a function of the `t Hooft
coupling $\lambda$.

The history of integrability methods in AdS/CFT can be traced back to the seminal paper \cite{Minahan:2002ve} published nearly ten years ago.
After very fast development in the field \cite{Bena:2003wd,Arutyunov:2004vx,Beisert:2005tm,Janik:2006dc,Beisert:2006ib,Beisert:2006ez} (a recent pedagogical review can be found in \cite{Beisert:2010jr}), a solution to the spectral problem
was soon obtained for asymptotically long single trace operators by means of the Bethe ansatz approach \cite{Beisert:2005fw}.
Soon after it was realized that the full spectrum
is governed by the Y-system equations \cite{Gromov:2009bc,Gromov:2009tv,Arutyunov:2009ur,Bombardelli:2009ns,Arutyunov:2009zu},
which also describe short operators. Moreover in \cite{Gromov:2009bc} the equations, suitable for the numerical studies,
were formulated for the $\sl_2$ operators of the type $\tr \, D^S Z^J$. The anomalous dimension
of the most famous among them --- the Konishi operator was found numerically in \cite{Gromov:2009zb}
as a function of $\lambda$ interpolating from the weak coupling expansion,
known explicitly from perturbative calculations in gauge theory up to four loops, to the strong coupling
string theory prediction \cite{Gubser:2002tv} known only to tree level.
Recently the sub-leading coefficient in the string theory expansion was found independently by three groups \cite{Gromov:2011de,Roiban:2011fe,Vallilo:2011fj}
and later considered in \cite{Beccaria:2011uz} using different variations
of one loop string quantization, confirming the numerical predictions of  \cite{Gromov:2009zb}. Even more recently a highly nontrivial observation was made in \cite{Basso:2011rs},
which allows one to reproduce the one loop result almost without any effort.

In this paper we use certain assumptions about the structure of the
strong coupling expansion of the scaling dimension for short operators, which makes it possible
to promote one loop results to next order. By using the one loop expression for a general $(S,J)$
folded string found in \cite{Frolov:2002av} and the conjecture found in \cite{Basso:2011rs}, we derive the second nontrivial strong coupling expansion
coefficient analytically. We then compare our results with the available numerical data from the TBA approach and find a rather promising agreement. We hope that the new FiNLIE\footnote{Finite set of Nonlinear Integral Equations}
 will lead to more precise tests of our results in the near future \cite{finlie}.

\section{Folded string}
The folded string is the strong coupling counterpart of the Wilson operators $\tr(D^S Z^J)$.
This class of operators in particular contains the Konishi operator that has been receiving a lot of attention recently.
\subsection{Tree level}
The classical energy of the folded string is a function of the Lorentz spin $S$,
twist $J$ and the mode number $n$. This function can be written in a parametric form
in terms of the branch points $a$ and $b$ \cite{Gromov:2011de,Beisert:2003ea,Kazakov:2004qf,Kazakov:2004nh}:
\beqa
\nn{2\pi {\cal S}}&=&\frac{ab+1}{ab}\[b E\(1-\tfrac{a^2}{b^2}\)-aK\(1-\frac{a^2}{b^2}\)\]\;,\\
{2\pi {\cal J}}&=&\frac{2\sqrt{(a^2-1)(b^2-1)}}{b}K\(1-\frac{a^2}{b^2}\)\;,\\
\nn{2\pi {\cal D}_{\rm tree}}&=&\frac{ab-1}{ab}\[b E\(1-\tfrac{a^2}{b^2}\)+aK\(1-\frac{a^2}{b^2}\)\]\;.
\eeqa
where ${\cal S},{\cal J},{\cal D}=\frac{S}{n\sqrt\lambda},\frac{J}{n\sqrt\lambda},\frac{\Delta}{n\sqrt\lambda}$.
In this paper we will concentrate on a special limit when $S$ is sent to zero.
In this limit one can write a more explicit expression for the square of the scaling dimension:
\beqa
 {\cal D}_{\rm tree}^2&=&{\cal J}^2+2 \, {\cal S} \, \sqrt{{\cal J}^2+1}+{\cal S}^2 \, \frac{2 {\cal J}^2+3}{2
   {\cal J}^2+2}-{\cal S}^3 \, \frac{{\cal J}^2+3}{8
   \left({\cal J}^2+1\right)^{5/2}}
   +{\cal O}\left({\cal S}^4\right)\;.
\eeqa
One can easily see that the coefficients in the expansion of ${\cal D}_{\rm tree}^2$
are considerably simpler than the same coefficients in the expansion of ${\cal D}_{\rm tree}$.

One can further notice \cite{Basso:2011rs} that the re-expansion of the function $\Delta^2$ in the large $\mu\equiv \lambda n^2$ limit with $S$ and $J$ fixed has a particularly nice structure
\beq
\Delta_{\rm tree}^2\!\!\!=\!J^2+S
\(
2\, \sqrt{\mu}+\frac{J^2}{\sqrt{\mu}}+\dots
\)
+S^2
\(
\frac{3}{2}-\frac{J^2}{2\mu}
+\dots
\)
-S^3
\(
\frac{3}{8\sqrt{\mu}}
-\frac{13 J^2}{16\sqrt[3]{\mu}}
+\dots
\)
+{\cal O}({ S}^4)
\label{dsquare_tree}
\eeq
where each next term in $S$
gets more and more suppressed for large $\lambda$.
This structure indicates that the expansion in large $\lambda$
and small $S$ should be easily computable, which is very important in the study of short operators. 
The structure in \eq{dsquare_tree} is a purely classical result. In the next section we discuss whether it is preserved when quantum corrections are taken into account.

\subsection{One loop}
Using the algebraic curve technique \cite{Gromov:2009zza,Gromov:2007aq,Gromov:2007ky,Gromov:2008ec,SchaferNameki:2010jy}
the result \eq{dsquare_tree} at one loop can be shown to be just a little bit more involved than the classical energy.
The derivation is described in \cite{Gromov:2011de} so we only quote the result here (see appendix \ref{appA} for more details).

Again, in the limit when ${\cal S}$ is sent to zero the result simplifies significantly.
Up to two orders in $\mathcal{S}$ we found the following expansion
\beqa
\label{delta_oneloop}
\Delta_{\rm 1-loop}&\simeq&
\frac{-{\cal S}}{2 \left({\cal J}^3+{\cal J}\right)}+{\cal S}^2\[\frac{3 {\cal J}^4+11 {\cal J}^2+17
   }{16 {\cal J}^3 \left({\cal J}^2+1\right)^{5/2}}
\!-\!\sum_{m>0,m\neq n}\frac{n^3m^2  \left(2 m^2+n^2 {\cal J}^2-n^2\right)}{{\cal J}^3 \left(m^2-n^2\right)^2
   \left(m^2+n^2 {\cal J}^2\right)^{3/2}}\]
   \;.
\eeqa
The next term in this expansion can be found in \eq{delta_oneloop_3}, \eq{delta_oneloop_4}. The sum is nothing but a sum over the fluctuation energies,
whereas the remaining terms originate from the ``zero"-modes
$m=n$, which have to be treated separately.
The sum can be very easily expanded for small ${\cal J}$.
It is easy to see that the expansion coefficients will be certain combinations
of zeta-functions. It is also easy to see that
the dependence on the mode number $n$ is rather nontrivial.

The expansion of the one loop energy first in small ${\cal S}$
up to a second order and then in small ${\cal J}$ reads
\beq
\label{delta_oneloop_sj}
\Delta_{\rm1-loop}\simeq
\left\{
\begin{array}{ll}
 -\frac{\mathcal{S}}{2 \mathcal{J}}+
 \mathcal{S}^2 \left(+\frac{1}{2\mathcal{J}^3}-\frac{3 \zeta_3}{2 \mathcal{J}}-\frac{1}{16 \mathcal{J}}\right) & \;\;,\;\;n=1 \\
 -\frac{\mathcal{S}}{2 \mathcal{J}}+
 \mathcal{S}^2 \left(+\frac{1}{2
   \mathcal{J}^3}-\frac{12 \zeta_3}{\mathcal{J}}-\frac{17}{16 \mathcal{J}}\right) & \;\;,\;\;n=2 \\
 -\frac{\mathcal{S}}{2 \mathcal{J}}+
 \mathcal{S}^2 \left(-\frac{5}{8
   \mathcal{J}^3}-\frac{81 \zeta_3}{2 \mathcal{J}}-\frac{7}{4 \mathcal{J}}\right) & \;\;,\;\;n=3 
\end{array}
\right.
\eeq
Expansions up to four orders in $\mathcal{S}$ and then in $\mathcal{J}$ are given in appendix \ref{AppS4}. We note that the contributions ${\cal S}^2/{\cal J}^3$
are universal for $n=1$ and $n=2$,
however starting from $n=3$ we get some nasty coefficient.
As we will discuss in the next section this could imply that
the naive generalization of the conjecture in \cite{Basso:2011rs} is not fully correct
for $n>2$. Also for $n=2$ we found  a similar anomaly at the order $S^3$.
\section{Discussion of the exact slope and its generalizations}
Let us take a close look at the conjecture in \cite{Basso:2011rs}. It says that
making expansions of the scaling dimension squared first in ${\cal S}\to 0$
and then in $\mu\to \infty$ should reveal the following structure
\beq\la{Delta}
\Delta^2=J^2+S
\(
A_1\sqrt{\mu}+A_2+\dots
\)
+S^2
\(
B_1+\frac{B_2}{\sqrt\mu}
+\dots
\)
+{\cal O}({ S}^3)\;,
\eeq
where the coefficients $A_i,\;B_i,\;C_i$ are some functions of $J$.
This is, as can be easily seen, a nontrivial constraint on $\Delta$ itself as
\beqa\la{cons}
\Delta&=&J+\frac{S}{2J}
\(
A_1\sqrt{\mu}+A_2+\frac{A_3}{\sqrt{\mu}}+\dots
\)\\
\nn&+&S^2
\(
- \frac{A_1^2}{8J^3} \, \mu
-  \frac{A_1A_2}{4J^3} \, \sqrt{\mu}
+\[\frac{B_1}{2J}-\frac{A_2^2+2A_1 A_3}{8J^3}\]
+
\[
\frac{B_2}{2J}
-\frac{A_2A_3+A_1A_4}{4J^3}
\]  \frac{1}{\sqrt\mu}
+\dots
\)
+{\cal O}(S^3)\;.
\eeqa
One of the results of \cite{Basso:2011rs}
is the exact formula for all the coefficients $A_i$. They can be found easily by expanding
a simple combination of Bessel functions, called the ``slope", around infinity and it produces~\cite{Basso:2011rs}:
\beq
A_1=2\;\;,\;\;
A_2=-1\;\;,\;\;
A_3=J^2-\frac{1}{4}\;\;,\;\;
A_4=J^2-\frac{1}{4}\dots\;.
\la{As}
\eeq
Comparing with our one-loop result we get\footnote{$B_2=-b$ in the notations of \cite{Basso:2011rs}.
The $-3\zeta_3$ term also arises in the formalism of \cite{Vallilo:2011fj} when formally extended to two loops.
A very similar $\zeta_3$ term can be also extracted from \cite{Roiban:2011fe}.
This gives extra support to our results.
We would like to thank L.Mazzucato and A.Tseytlin for pointing this out.
}
\beq\la{BB}
B_1=\frac{3}{2}\;\;,\;\;
B_2=
\left\{
\bea{ll}
-3\,\zeta_3+\frac{3}{8}&\;\;,\;\;n=1\\
-24\,\zeta_3-\frac{13}{8}&\;\;,\;\;n=2\\
-81\,\zeta_3-\frac{24}{8}&\;\;,\;\;n=3
\eea
\right.\;.
\eeq
We should, however, notice that for $n>1$ we were not able to fully satisfy \eq{cons}.
One example is the coefficient in front of $S^2/J^3$, which for $n=3$ is $-5/8$, whereas \eq{cons} predicts $1/2$. We observe that only for $S^2$, $S^3$ and higher order terms do we find such disagreements and it is interesting to note that the coefficients for $S$ order terms seem to be correct for any $n$\footnote{We indeed verified numerically that the naive replacement $\lambda\to n^2\lambda$ works at weak coupling at least to two loops.}.
These observations imply that the generalization of the original slope function,
which is done by a naive replacement $\lambda\to n^2\lambda$, is not correct for the cases when $n>1$ and thus either the coefficients in \eq{As} or the conjecture itself should be modified to accommodate this.
We discuss this in details in the next section \ref{inconsistencies}.

\subsection{Inconsistencies in the next orders}\la{inconsistencies}

The analysis in the previous sections was done only up to second order in the small $S$ expansion. The appendix \ref{AppS4} contains our result for the
one-loop quantization of the $n$-times folded string up to the order $S^4$. For $n=1$ our result is in perfect agreement with the conjectured structure \eq{Delta}, yet for cases with $n>1$ there are inconsistencies.
For $n=2$ the first inconsistency appears in the $\frac{S^3\mu}{J^4}$ term and for $n=3$ there are already inconsistencies at order $S^2$. We found that for $n>1$ one has to modify the structure in \eq{Delta} by
including negative coefficients in order for it to be consistent with our one-loop results. E.g. for $n=2$ the structure has to be modified starting with the $S^3$ term, which now becomes
\beq
\(
C_{-2}\;\mu+\frac{C_1}{\sqrt\mu}+\frac{C_2}{\mu}+\dots
\) S^3
\eeq
with $C_{-2}=\frac{12}{J^4}$. To the next order in $S$ we find
\beq
\(
D_{-4}\;\mu^{3/2} +D_{-2}\;\sqrt{\mu}+ \frac{D_{0}}{\sqrt\mu}+\frac{D_1}{\mu}+\dots
\) S^4
\eeq
where $D_{-4}=-\frac{78}{J^6},\;D_{-2}=-\frac{36}{J^4},\;D_0=\frac{21}{2J^2}$.

For $n=3$ the first modification already occurs at order $S^2$ and it can be resolved if the term $-\frac{9 S^2\sqrt\mu}{4 J^2}$ is added to \eq{Delta}.
Thus effectively the conjectured structure \eq{Delta} has to be modified as in the $n=2$ case by including negative coefficients, which now depend on $n$ in a nontrivial way.
It is also worth noticing that since inconsistencies start appearing at orders of $\frac{S^2}{J^2}$ and $\frac{S^3}{J^4}$ for $n=3$ and $n=2$ respectively, one might guess that
there should be an inconsistency at order $\frac{S^4}{J^6}$ for $n=1$, however we found no such thing.

This study of inconsistencies reveals that the proposed modifications to the structure of \eq{Delta} have growing powers of $\mu$,
thus one should resum them together with similar singular terms which may arise in higher loop levels before being able to make justified predictions
for short operators ($S\sim J\sim 1$) at strong coupling when $n>1$.

\section{Two loop prediction}
The equation \eq{Delta} allows one to
make a very nontrivial prediction for the strong coupling
expansion of operators with fixed length $J$ and
the number of derivatives $S$. For that end we simply fix $S$ and $J$ in \eq{Delta}
and expand for large $\lambda$ or, equivalently, $\mu$. This procedure gives:
\beq
\Delta_{S,J,n}\simeq\sqrt{2S}\mu^{1/4}
+\frac{2J^2+3S^2-2S}{4\,(2 S)^{1/2}\,\mu^{1/4}}
+\frac{-21S^4+(32B_2+12)S^3+(20J^2-12)S^2+8J^2 S-4J^4}{32\,(2S)^{3/2}\,\mu^{3/4}}\;
\eeq
where $B_2$ is given in \eq{BB}.
Note that according to our observations there are some inconsistencies in the conjecture that this derivation relies on when $n>1$ and thus this result should be treated with great care.\footnote{We assume that the results of \cite{Basso:2011rs} for the slope function
can be lifted by generalizing with the simple replacement $\lambda\to n^2\lambda$ when $n>1$. 
We indeed verified this numerically with high precision at weak coupling up to two loops and this is also in agreement
with our one loop strong coupling results. I.e. the slope function and hence the coefficients $A_i$ in \eq{As} are still correct after the replacement, but as argued before, the structure of the expansion \eq{Delta} may need to be modified.}

Let us write the result more explicitly for a particular important case of two magnons
\beq
\Delta_{2,J,1}=2 \, \lambda^{1/4}+
\frac{\frac{J^2}{4}+1}{\lambda^{1/4}}+\frac{-\frac{J^4}{64}+\frac{3 J^2}{8}-3\, \zeta
   (3)-\frac{3}{4}}{\lambda^{3/4}}\;.
\eeq
In the next section we compare our prediction with the available TBA data.

\section{TBA numerics}

\input{TBA}

\section{Conclusions}
In this letter we made a prediction for the two loop
coefficient for some short $\sl(2)$ operators, including the Konishi operator,
and compared the result to existing TBA data.
Our results seem to agree well. Nevertheless, it is
very important to get better precision for the anomalous dimensions
of short operators.
This should be possible to do using the novel FiNLIE approach developed in \cite{finlie}.
It would also be interesting to check our results with operators having $S>2$, since it is known how the FiNLIE equations should look like for such high twist cases.\footnote{Recently the case with operators having $S=3$ in the $\su(2)$ sector was considered in \cite{Arutyunov:2011mk} using an infinite system of equations.}
Finally it is also very important to accomplish the same result from a first principle calculation, recent progress in that direction can be seen in \cite{Passerini:2010xc}.

\appendix

\section*{Acknowledgements}
We would like to thank S.~Frolov and A.~Tseytlin for discussions (also special thanks to S.~Frolov for sharing his TBA data), P.~Vieira for stimulating conversations and
especially B.~Basso for very evaluating comments.

{\bf Note added}: while the current version of the paper was under consideration in JHEP
the paper \cite{Frolov:2012zv} was published containing similar conclusions
about the numerics for n=1 and n=2 states.

\newpage
\section{Exact formulae for one-loop correction}\la{appA}
\subsection{Main formula for one-loop correction and notations}
In \cite{Gromov:2011de} a general formula was derived describing the one loop correction
to the energy of the generic $(S,J,n)$ folded string solution.
There are three contributions to one loop energy shift that are different by their nature.
They can be separated into an ``anomaly" contribution, a contribution from the dressing phase
and a wrapping contribution, which is missing in the ABA approach, but present in the
Y-system
\beq
\Delta_{\rm1-loop}=\delta \Delta_{\rm anomaly} +\delta \Delta_{\rm dressing} + \delta \Delta_{\rm wrapping}\;,
\eeq
where each of these contributions is simply an integral of some closed form expression,
\beqa
\label{eq:delta_E_3}
  \delta \Delta_{\rm anomaly} &=& -\frac{4}{ab-1}\int_a^b \frac{dx} {2\pi i}
  \frac{y(x)}{x^2-1} \partial_x \log \sin p_{\hat 2}\;,\\
  \label{eq:delta_E_1}
  \delta \Delta_{\rm dressing} &=& \sum_{ij} (-1)^{F_{ij}} \int\limits_{-1}^{1} \frac{dz} {2\pi
    i} \left( \Omega^{ij}(z) \, \partial_z \frac{i (p_i -p_j)}{2} \right)\;,\\
  \label{eq:delta_E_2}
  \delta \Delta_{\rm wrapping} &=& \sum_{ij} (-1)^{F_{ij}} \int\limits_{-1}^{1} \frac{dz} {2\pi
    i} \left( \Omega^{ij}(z) \, \partial_z \log (1- e^{-i(p_i -p_j)}) \right)\;.
\eeqa
in this sum $i$ takes values $\hat 1,\hat 2,\tilde 1,\tilde 2$ whereas $j$ runs over
$\hat 3,\hat 4,\tilde 3,\tilde 4$.

Let us explain the notations. The quasi-momenta:
\beqa
  \label{eq:p_a}\nn
  p_{\hat 2} &=& \pi n - 2\pi n{\cal J} \left( \frac{a}{a^2-1} -
    \frac{x}{x^2-1} \right) \sqrt{\frac{(a^2-1)
      (b^2-x^2)}{(b^2-1)(a^2-x^2)}} \\ \nn &+& \frac{8\pi n a b {\cal S} F_1(x)}{(b-a)(ab+1)} +
  \frac{2\pi n {\cal J} (a-b) F_2(x)}{\sqrt{(a^2-1)(b^2-1)}},\\
  \label{eq:p_s}
  p_{\tilde 2} &=& \frac{2\pi {\cal J}x}{x^2-1}.
\eeqa
The integer $n$ (the mode number) is related to the number of spikes.
All the other quasi-momenta can be
found from
\beqa
  \label{eq:quasimomenta_symmetry_A}
  p_{\hat{2}} (x) &=& -p_{\hat{3}}(x) = -p_{\hat{1}}(1/x) = p_{\hat{4}}
  (1/x)\;,\\
  \label{eq:quasimomenta_symmetry_S}
  p_{\tilde{2}} (x) &=& -p_{\tilde{3}} (x) = p_{\tilde{1}} (x) =
  -p_{\tilde{4}}(x)\;.
\eeqa
The functions $F_1(x)$ and $F_2(x)$ can be expressed
in terms of the elliptic integrals:
\begin{eqnarray}
 \label{eq:not}
 F_1(x) &=& i F \left( i \sinh^{-1}
 \sqrt{\frac{(b-a)(a-x)}{(b+a)(a+x)}} | \frac{(a+b)^2}{(a-b)^2} \nonumber
 \right)\;, \\
\nn
 F_2(x) &=& i E \left( i \sinh^{-1}
 \sqrt{\frac{(b-a)(a-x)}{(b+a)(a+x)}} | \frac{(a+b)^2}{(a-b)^2}
 \right)\;.
 \end{eqnarray}
Finally the off-shell fluctuation energies are
\begin{eqnarray}\nn
  \label{eq:frequencies}
  \Omega^{\hat{1}\hat{4}} (x) &=& -\Omega^{\hat{2}\hat{3}}(1/x) -2\;,\;\;\\
  \Omega^{\hat{1}\hat{3}} (x) &=& \Omega^{\hat{2} \hat{4}} (x) = \frac12
  \Omega^{\hat{1}\hat{4}}(x) + \frac12 \Omega^{\hat{2}\hat{3}}(x)\;,\nn\\
  \Omega^{\hat{1}
    \tilde{3}}(x)
&=&
  \Omega^{\hat{1}\tilde{4}}(x)
    =  \Omega^{\hat{4}\tilde{1}}(x)
    =  \Omega^{\hat{4}\tilde{2}}(x)
     =
  \frac12\, \Omega^{\tilde{2}\tilde{3}} (x) + \frac12\,\Omega^{\hat{1}\hat{4}}(x),\\
  \Omega^{\hat{2}\tilde{3}}(x)
&=&
  \Omega^{\hat{2}\tilde{4}}(x)
=
  \Omega^{\tilde{1}\hat{3}}(x)
=
  \Omega^{\tilde{2}\hat{3}}(x)
= \frac12\, \Omega^{\tilde{2}\tilde{3}}(x) + \frac12\, \Omega^{\hat{2}\hat{3}}(x),\nn\\
  \Omega^{\tilde{2}\tilde{3}}(x)
&=&
  \Omega^{\tilde{2}\tilde{4}}(x)
=
  \Omega^{\tilde{1}\tilde{3}}(x)
=
  \Omega^{\tilde{1}\tilde{4}}(x)\;,\nn
\end{eqnarray}
where
\beqa
\label{oma}
\Omega^{\tilde 2\tilde 3}(x)&=&\frac{2}{a
  b-1}\frac{\sqrt{a^2-1}\sqrt{b^2-1}}{x^2-1}\;,\\
\label{oms}
\Omega^{\hat 2\hat 3}(x)&=&\frac{2}{a b-1}\(1-\frac{y(x)}{x^2-1}\)\;.
\eeqa
and $y(x)=\sqrt{x-a} \sqrt{a+x} \sqrt{x-b} \sqrt{b+x}$.

In the small $\cal{S}$ limit these expressions can be expanded,
\beqa\label{anomaly_dressing}
	&&\delta\Delta_{anomaly} = \frac{-1}{2( \mathcal{J}^3 + \mathcal{J})} \, \mathcal{S}
	+ \left[ \frac{2 \, \mathcal{J}^4 + 15 \, \mathcal{J}^2 + 4}{16 \, \mathcal{J}^3(\mathcal{J}^2 + 1)^{5/2}} - \frac{\pi^2 n^2}{12 \, \mathcal{J}^3 \, \sqrt{\mathcal{J}^2 + 1}} \right] \mathcal{S}^2   \\
	&&+ \left[ \frac{3 \, \mathcal{J}^8 -32 \, \mathcal{J}^6 - 146 \, \mathcal{J}^4 - 68 \, \mathcal{J}^2 - 16}{64 \, \mathcal{J}^5 (1+ \, \mathcal{J}^2)^4} + \frac{\pi^2 n^2 (\, \mathcal{J}^4 + 4 \, \mathcal{J}^2 + 2)}{24 \, \mathcal{J}^5 (1 + \, \mathcal{J}^2)^2} + \frac{\pi^4 n^4}{180 \, \mathcal{J}^5} \right] \mathcal{S}^3 \nonumber
	+ \mathcal{O}(\mathcal{S}^4)  \\
	&&\delta\Delta_{dressing} = \left[ \frac{n \, (\mathcal{J}^2 + 2) \, \mathrm{coth}^{-1} (\sqrt{\, \mathcal{J}^2 + 1} + \, \mathcal{J})}{\, \mathcal{J}^3(\, \mathcal{J}^2 + 1)^{3/2}} -\frac{n}{2 \, \mathcal{J}^3 (\, \mathcal{J}^2 + 1)}  \right] \mathcal{S}^2 \nonumber \\
	&&+ \left[ -\frac{n(3 \, \mathcal{J}^6 + 13 \, \mathcal{J}^4 + 22 \, \mathcal{J}^2 + 8)\, \mathrm{coth}^{-1} (\sqrt{\, \mathcal{J}^2 + 1} + \, \mathcal{J})}{2 \, \mathcal{J}^5 (1 + \, \mathcal{J}^2)^3} + \frac{n(9 \, \mathcal{J}^4 + 31 \, \mathcal{J}^2 + 10)}{12 \, \mathcal{J}^5 (1 + \, \mathcal{J}^2)^{5/2}} \right] \mathcal{S}^3 \nonumber
	+ \mathcal{O}(\mathcal{S}^4)
\eeqa
the expansion of the third integral $\delta\Delta_{wrapping}$ is more complicated, and we advice the reader to use
the equation \eq{delta_oneloop} instead which includes all contributions. What we can, however, say is that
	$\delta\Delta_{wrapping} = {\cal O}(e^{-2\pi{\cal J}})$
and thus this term is irrelevant for the large ${\cal J}$ expansion.
This makes the expressions \eq{anomaly_dressing} particularly convenient for
small ${\cal S}$ followed by large ${\cal J}$ expansions,
where as the exact ${\cal J}$ expression in \eq{delta_oneloop} is not
very convenient since the sum of the expansion does not converge.

\newpage
\subsection{One Loop $(S,J)$ Folded String {\it Mathematica} {Code}}
In order to fix all our conventions as well as for the convenience of the
reader we include a simplified  {\it Mathematica} {code}
we used to check our results numerically\\
{\footnotesize
\verb" "\\
\verb"GS=((2*(a*b+1))*(b*EllipticE[1-a^2/b^2]-a*EllipticK[1-a^2/b^2]))/(4*Pi*a*b);"\\
\verb"GJ=((4*Sqrt[(a^2-1)*(b^2-1)])*EllipticK[1-a^2/b^2])/(4*Pi*b);"\\
\verb"y=Sqrt[x-a]*Sqrt[x+a]*Sqrt[x-b]*Sqrt[x+b];"\\
\verb"F1[x_] =I*EllipticF[I*ArcSinh[Sqrt[-(((a-b)*(a-x))/((a+b)*(a+x)))]], (a+b)^2/(a-b)^2];"\\
\verb"F2[x_] =I*EllipticE[I*ArcSinh[Sqrt[-(((a-b)*(a-x))/((a+b)*(a+x)))]], (a+b)^2/(a-b)^2];"\\
\verb"pA[x_] =n*Pi-2*Pi*n*j*(a/(a^2-1)-x/(x^2-1))*Sqrt[((a^2-1)*(b^2-x^2))/((b^2-1)*(a^2-x^2))] +"\\
\verb"       (8*a*b*s*Pi*n*F1[x])/((b-a)*(a*b+1))+(2*Pi*n*j*(a-b)*F2[x])/Sqrt[(a^2-1)*(b^2-1)];"\\
\verb"pS[x_]=(2*Pi*n*j*x)/(x^2-1);"\\
\verb"X[z_]=z+Sqrt[z^2-1];"\\
\verb"OA[x_]=(2*(1-y/(x^2-1)))/(a*b-1);"\\
\verb"OS[x_]=(2*(-(y /. x->1)))/((a*b-1)*(x^2-1));"\\
\verb"ab[j_, s_] :=ab[j, s]=Chop[FindRoot[{s==GS, j==GJ}, {{b, Sqrt[j^2+1]+j+Sqrt[s]/10}"\\
\verb"                                                    ,{a, Sqrt[j^2+1]+j-Sqrt[s]/10}}]];"\\
\verb"OneLoop[jj_, ss_, nn_] := Block[{sb0=Join[ab[jj, ss], {j->jj, s->ss, n->nn}]},"\\
\verb"tn0=(2*Im[pA[X[z]]-pS[X[z]]]*Im[D[OA[X[z]]-OS[X[z]], z]])/Pi /. sb0;"\\
\verb"Edressing=NIntegrate[tn0, {z, 0, 1}];"\\
\verb"tn1=(2*D[OS[X[z]], z]*Log[((1-Exp[(-I)*pS[X[z]]-I*pA[X[z]]])*(1-Exp[(-I)*pS[X[z]]+I*pA[1/X[z]]]))/"\\
\verb"                                                              (1-Exp[-2*I*pS[X[z]]])^2])/Pi /. sb0;"\\
\verb"tn2=-((2*D[OA[X[z]], z]*Log[((1-Exp[-2*I*pA[X[z]]])*(1-Exp[(-I)*pA[X[z]]+I*pA[1/X[z]]]))/"\\
\verb"                                                  (1-Exp[(-I)*pS[X[z]]-I*pA[X[z]]])^2])/Pi) /. sb0;"\\
\verb"Ewrapping=NIntegrate[Im[tn1+tn2], {z, 0, 1}];"\\
\verb"tn=-((4*y*D[Log[Sin[pA[x]]], x])/((a*b-1)*(2*Pi*I)*(x^2-1))) /. sb0;"\\
\verb"Eanomaly=Re[NIntegrate[tn, {x, a /. sb0, ((a+b)*(1+I))/(2*10) /. sb0, b /. sb0}]];"\\
\verb"Edressing+Ewrapping+Eanomaly];"
\verb" "\\
}\\
To compute $\Delta_{\rm 1-loop}$ simply run \verb"OneLoop["${\cal J},{\cal S}, n$\verb"]" in {\it Mathematica}.

\section{$S^3$ and $S^4$ order}\la{AppS4}
The $\mathcal{S}^3$ order term in the expression of \eq{delta_oneloop} is given by
\beqa
\label{delta_oneloop_3}
\delta\Delta_{1-loop}^{(3)} &=&-\frac{6 \, \mathcal{J}^8 + 48 \, \mathcal{J}^6+ 138 \, \mathcal{J}^4 + 352 \, \mathcal{J}^2 + 117 }{64 \, \mathcal{J}^5 (\, \mathcal{J}^2 + 1)^4 }  \\
 &+& \sum_{m>0,m\neq n}
\nn \frac{P_3(n,m,{\cal J}) }{2 \, \mathcal{J}^5 (\, \mathcal{J}^2 + 1)^{3/2} (m^2 - n^2)^4  (\mathcal{J}^2 n^2 + m^2)^{5/2}}
\eeqa
and the $\mathcal{S}^4$ order term is given by
\beqa
\label{delta_oneloop_4}
\delta\Delta_{1-loop}^{(4)} &=&
\frac{45 {\cal J}^{12}+717 {\cal J}^{10}+3429 {\cal J}^8+11205
   {\cal J}^6+27601 {\cal J}^4+15789 {\cal J}^2+3305}{1024 {\cal J}^7
   \left({\cal J}^2+1\right)^{11/2}} \\
   &-& \sum_{m>0,m\neq n}\frac{P_4(n,m,\cal J)}{16 {\cal J}^7 \left({\cal J}^2+1\right)^3 (m^2-n^2)^6 \left(m^2+n^2
   {\cal J}^2\right)^{7/2}} \nonumber
\eeqa
where
\beqa
P_3(n,m,{\cal J})=&+&
m^{10} n^3 \left(4 {\cal J}^4+11 {\cal J}^2+6\right)+2 m^8 n^5 \left(3
   {\cal J}^6+5 {\cal J}^4-6 {\cal J}^2-6\right)\\
\nn&+&2 m^6 n^7 \left({\cal J}^8-4 {\cal J}^6-11 {\cal J}^4+6 {\cal J}^2+9\right)+2 m^4
   n^9 \left(-2 {\cal J}^8+9 {\cal J}^6+29 {\cal J}^4+14
   {\cal J}^2-2\right)\\&+&m^2 n^{11} {\cal J}^2 \left(10 {\cal J}^6+16
\nn   {\cal J}^4-2 {\cal J}^2-7\right)\;,
\\P_4(n,m,{\cal J})=
&+&4 m^{16} n^3 \left(8 {\cal J}^8+42 {\cal J}^6+85 {\cal J}^4+68
   {\cal J}^2+20\right)
\\&+&m^{14} n^5 \left(80 {\cal J}^{10}+302 {\cal J}^8+199
   {\cal J}^6-703 {\cal J}^4-936 {\cal J}^2-340\right)
\nn\\&+&m^{12} n^7 \left(64
   {\cal J}^{12}-12 {\cal J}^{10}-893 {\cal J}^8-1765 {\cal J}^6+151
   {\cal J}^4+1587 {\cal J}^2+740\right)
\nn\\&+&m^{10} n^9 \left(16 {\cal J}^{14}-222
   {\cal J}^{12}-587 {\cal J}^{10}+1209 {\cal J}^8+5444 {\cal J}^6+4374
   {\cal J}^4+388 {\cal J}^2-520\right)
\nn\\&+&2 m^8 n^{11} \left(-38 {\cal J}^{14}+200
   {\cal J}^{12}+1446 {\cal J}^{10}+2505 {\cal J}^8+769 {\cal J}^6-511
   {\cal J}^4+17 {\cal J}^2+210\right)
\nn\\&+&m^6 n^{13} \left(200 {\cal J}^{14}+572
   {\cal J}^{12}+206 {\cal J}^{10}-176 {\cal J}^8+2199 {\cal J}^6+3085
   {\cal J}^4+1068 {\cal J}^2-60\right)
\nn\\&+&m^4 n^{15} {\cal J}^2 \left(-76
   {\cal J}^{12}+464 {\cal J}^{10}+2920 {\cal J}^8+5315 {\cal J}^6+3667
   {\cal J}^4+643 {\cal J}^2-173\right)
\nn\\&+&m^2 n^{17} {\cal J}^4 \left(256
   {\cal J}^{10}+962 {\cal J}^8+1221 {\cal J}^6+401 {\cal J}^4-250
   {\cal J}^2-148\right)\;.\nn
\eeqa
The expansion \eq{delta_oneloop_sj} can also be written in higher orders of $\mathcal{S}$ and $\mathcal{J}$, for $n=1$ we get
{\small
\beqa
	&&\Delta_{1-loop} = \( \frac{-1}{2\, \mathcal{J}} + \frac{\mathcal{J}}{2} \) \, \mathcal{S} + \( \frac{1}{2 \, \mathcal{J}^3} - \[ \frac{3 \, \zeta_3}{2} + \frac{1}{16} \] \frac{1}{ \mathcal{J}} + \[\frac{3 \, \zeta_3}{2} + \frac{15 \, \zeta_5}{8} - \frac{21}{32} \] \, \mathcal{J}  \) \, \mathcal{S}^2  \\
	&+& \( \frac{-3}{4 \, \mathcal{J}^5} + \[ \frac{3 \, \zeta_3}{2} + \frac{3}{16} \] \frac{1}{ \mathcal{J}^3} + \[ \frac{9 \, \zeta_3}{8} - \frac{1}{32} \] \frac{1}{\mathcal{J}} + \[ \frac{5}{4} - \frac{17 \, \zeta_3}{4} - \frac{65 \, \zeta_5}{16} - \frac{35 \, \zeta_7}{16} \] \, \mathcal{J}  \) \, \mathcal{S}^3\nonumber
\\  &+&\(
\frac{5}{4 {\cal J}^7}-
\[\frac{7}{32}+\frac{9 \zeta_3}{4}\]\frac{1}{{\cal J}^5}
-\[{\frac{3 \zeta_3}{4}-\frac{15 \zeta_5}{16}+\frac{5}{32}}\]\frac{1}{{\cal J}^3}-
\[{\frac{145 \zeta_3}{64}+\frac{45 \zeta_5}{32}+\frac{175 \zeta_7}{128}+\frac{27}{1024}}\]\frac{1}{{\cal J}}
   \){\cal S}^4 + \mathcal{O}(\mathcal{S}^5)\;,\nn
\eeqa}
for $n=2$,
{\small
\beqa
	&&\Delta_{1-loop} = \( \frac{-1}{2\, \mathcal{J}} + \frac{\mathcal{J}}{2} \) \, \mathcal{S} + \( \frac{1}{2 \, \mathcal{J}^3} - \[ 12 \, \zeta_3 + \frac{17}{16} \] \frac{1}{ \mathcal{J}} + \[12 \, \zeta_3 + 60 \, \zeta_5 + \frac{27}{32} \] \, \mathcal{J}  \) \, \mathcal{S}^2   \\
	&+& \( \frac{21}{4 \, \mathcal{J}^5} + \[ 12 \, \zeta_3 + \frac{19}{16} \] \frac{1}{ \mathcal{J}^3} + \[ 9 \, \zeta_3 + \frac{47}{32} \] \frac{1}{\mathcal{J}} - \[ \frac{19}{4} + 34 \, \zeta_3 + 130 \, \zeta_5 + 280 \, \zeta_7 \] \, \mathcal{J}  \) \, \mathcal{S}^3 \nonumber\\
  &+&\(
-\frac{175}{4 {\cal J}^7}-
\[\frac{727}{32}+18 \zeta_3\]\frac{1}{{\cal J}^5}
-\[{{6 \zeta_3}{}-30 \zeta_5-\frac{155}{32}}\]\frac{1}{{\cal J}^3}-
\[{\frac{145 \zeta_3}{8}+{45 \zeta_5}+{175 \zeta_7}+\frac{7419}{1024}}\]\frac{1}{{\cal J}}
   \){\cal S}^4 + \mathcal{O}(\mathcal{S}^5)\;,\nn
\eeqa}
and finally for $n=3$,
\beqa
	&&\Delta_{1-loop} = \( \frac{-1}{2\, \mathcal{J}} + \frac{\mathcal{J}}{2} \) \, \mathcal{S} + \( \frac{-5}{8 \, \mathcal{J}^3} - \[ \frac{81 \, \zeta_3}{2} + \frac{7}{4} \] \frac{1}{ \mathcal{J}} + \[\frac{81 \, \zeta_3}{2} + \frac{3645 \, \zeta_5}{8} - \frac{147}{64} \] \, \mathcal{J}  \) \, \mathcal{S}^2  \\
	&+& \( \frac{1245}{32 \, \mathcal{J}^5} + \[ \frac{81 \, \zeta_3}{2} + \frac{39}{16} \] \frac{1}{ \mathcal{J}^3} + \[ \frac{243 \, \zeta_3}{8} + \frac{89}{32} \] \frac{1}{\mathcal{J}} - \[ \frac{89}{8} + \frac{459 \, \zeta_3}{4} + \frac{15795 \, \zeta_5}{16} + \frac{76545 \, \zeta_7}{16} \] \, \mathcal{J}  \) \, \mathcal{S}^3 \nonumber \\
	&-& \( \frac{258785}{512 \, \mathcal{J} ^7} + \[ \frac{243 \, \zeta_3}{4} + \frac{251423}{1024} \] \frac{1}{\mathcal{J}^5} - \[ \frac{3645 \, \zeta_5}{16} - \frac{81 \, \zeta_3}{4} + \frac{256229}{4096} \] \frac{1}{\mathcal{J} ^3} \right.  \nonumber \\
	&+& \left. \frac{27 \, (907200 \, \zeta_7 + 103680 \, \zeta_5 + 18560 \, \zeta_3 + 13457)}{8192 \, \mathcal{J} } \) \, \mathcal{S}^4 + \mathcal{O}(\mathcal{S}^5). \nonumber
\eeqa

\newpage
\bibliography{bibliography}

\end{document}

%% file: TBA.tex
\begin{figure}[h]
\begin{overpic}[width=0.99\textwidth]{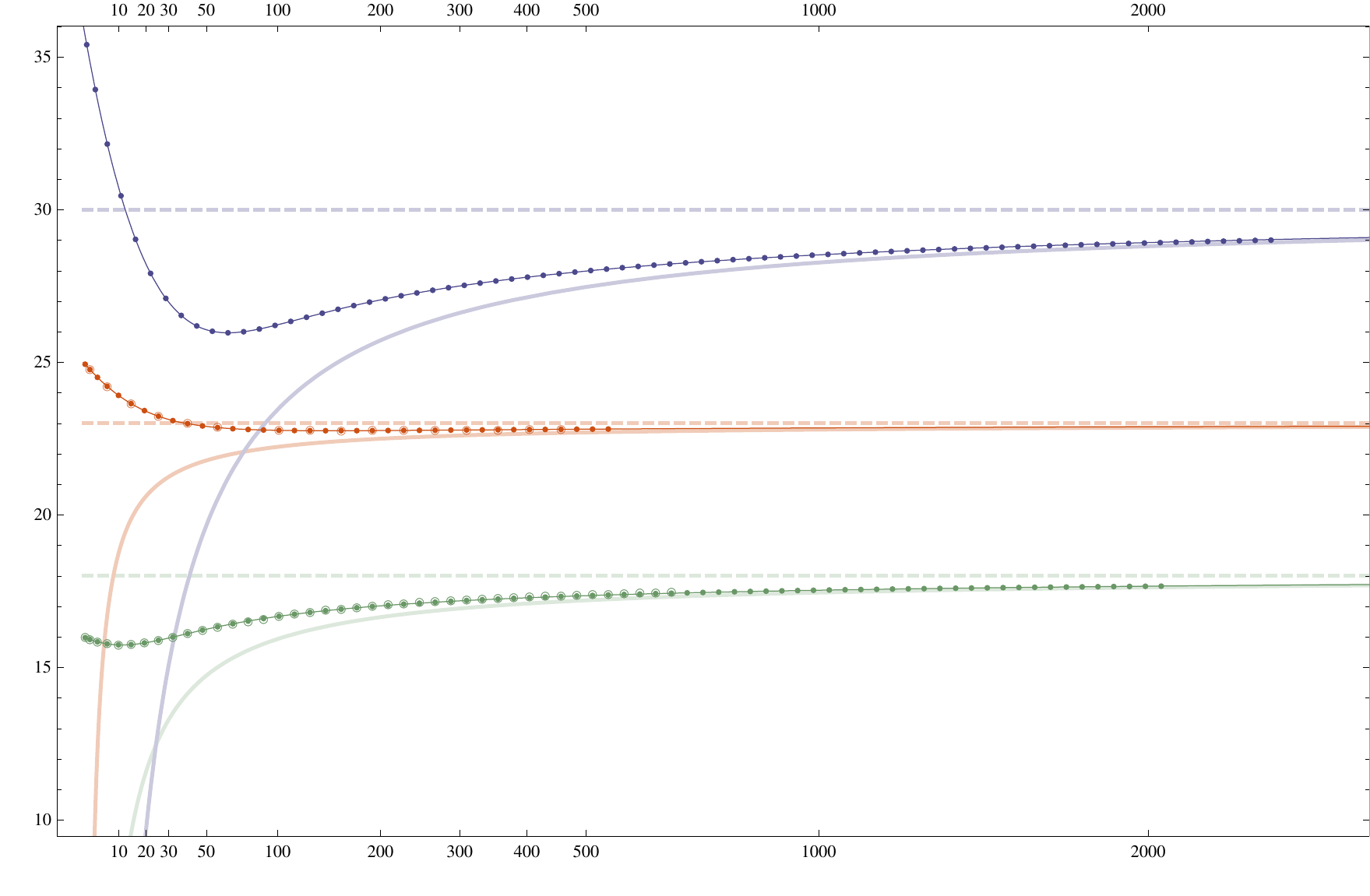}
\put(300,109){$S=2, J=2, n=1$}
\put(300,159){$S=2, J=3, n=1$}
\put(300,225){$S=2, J=4, n=2$}
\put(220,0){$\lambda$}
\put(-5,120){\begin{sideways}$\Delta^2 - y(n^2\lambda)$\end{sideways}}
\end{overpic}
\caption{Comparison of numeric TBA data to analytic predictions and fits. The plot shows the dependence of the scaling dimension squared of various operators on the coupling $\lambda$ with the leading order contributions subtracted. Solid dots represent numerical data taken from \cite{Frolov:2010wt}, empty circles for the Konishi state are taken from \cite{Gromov:2009zb} and empty circles for the $J=3$ state are numerics from \cite{Gromov:2011de}. $S=2, J=4, n=2$ points are from \cite{frolov_private}. Solid lines represent fits and the opaque solid lines of corresponding colors show our predictions. Dashed lines stand for $\lambda^{-1/4}$ predictions.}
\label{fig:frolov_compare}
\end{figure}
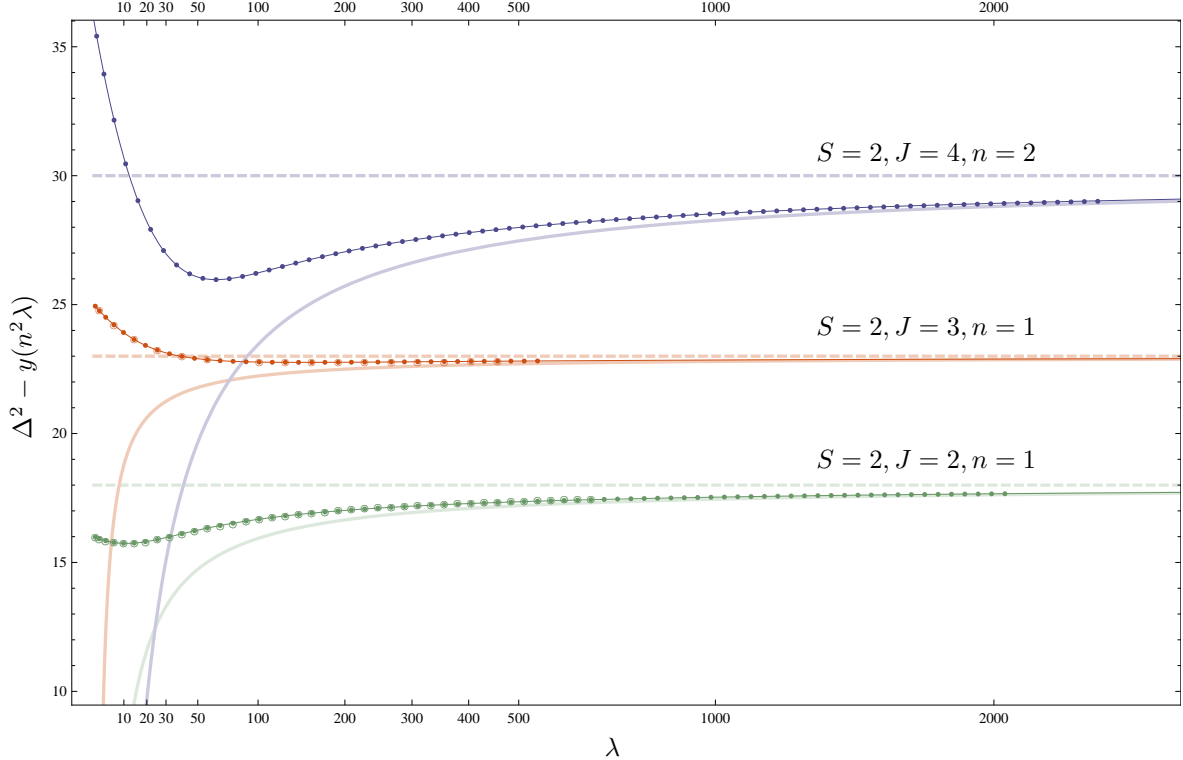

In order to extract strong coupling asymptotics from available TBA data, we performed numerical fits of Pad\'{e} type. First we changed variables from $\lambda$ to 
\begin{equation}
	y(\lambda) = \sqrt{\lambda} \frac{\partial}{\partial\sqrt{\lambda}} \log I_2(\sqrt{\lambda}) - 2,
\end{equation}
which seems arbitrary, but nevertheless is convenient because scaling dimension dependence on $y$ looks nearly linear and automatically captures some important analytical features. We then represent the scaling dimension as the square root of a rational function of two polynomials in $y$ with some of the unknown coefficients chosen so as to fix the leading order weak and strong coupling behaviours. So for example, for the Konishi operator we chose
\begin{equation*}
	\Delta_{2,2,1} = \sqrt{18 + 4y + \frac{-2 + \sum_{i=1}^{P} a_i y^i}{1 + \sum_{i=1}^{P+1} b_i y^i}},
\end{equation*}
because one can easily verify that the weak coupling expansion of this function is given by
\begin{equation*}
	\Delta_{2,2,1} = 4 + \mathcal{O}(g^2),
\end{equation*}
and the strong coupling expansion is given by
\begin{equation*}
	\Delta_{2,2,1} = 2 \lambda^{1/4} + \frac{2}{\lambda^{1/4}} + \mathcal{O}(\lambda^{-3/4}).
\end{equation*}
This way the leading order behaviour is fixed and next to leading order coefficients are combinations of the unknowns $a_i$ and $b_i$, which we then find by the method of least squares. The number of fit coefficients $P$ is chosen so that their values after fitting would be of order one, which would imply that the fit is reasonable. Though the procedure seems ad hoc, it produces incredibly good fits, which agree very well with both weak and strong coupling expansions. Fits to available TBA numerical data are shown in Fig. \ref{fig:frolov_compare}, where dots represent numerical values and the solid lines are our fits\footnote{For some of the fits we took the first 50 points from the corresponding data set, since we suspected the precission to be lower for higher values of $\lambda$. Also, these points were enough to get stable fits.}. Expanding our fits in powers of $\lambda$ at strong coupling we were able to compare the $\lambda^{-3/4}$ coefficients in the expansions to our predictions. These are summarized in Table \ref{tab:coefficients} for various operators. We see that our predictions agree with numerical data very well. The table also lists the weak coupling expansion coefficients of $g^2$ (tree level is fixed by hand), which agree with remarkable precision to Bethe ansatz predictions, once again indicating that the fits work well in both ends of the coupling range.

\begin{table}[t]
\begin{tabular}{|l||rl|l|l||l|l|l|}
  \hline
  $(S,J,n)$ & \multicolumn{2}{|l|}{$(n^2 \lambda)^{-3/4}$ prediction} & $(n^2 \lambda)^{-3/4}$ fit & error & $g^2$ analytical & $g^2$ fit & fit order\\
  \hline
  $(2,2,1)$ & $1/2 - \zeta_3 $&$= -3.1062$ & $-3.0739$ & $1.0\%$ & $12$ & $12.0108$ & 6\\
  $(2,3,1)$ & $87/64 - 3\,\zeta_3 $&$= -2.2468$ & $-2.2296$ & $0.8\%$ & $8$ & $8.0039$ & 5 \\
  $(2,4,2)$ & $-3/4 -24 \, \zeta_3 $&$= -29.5994$ & $-30.0547$ & $1.5\%$ & $14.4721$ & $14.4428$ & 5\\
  \hline
\end{tabular}
\caption{Comparisons of strong coupling expansion coefficients for $\lambda^{-3/4}$ obtained from fits to TBA data versus our predictions for various operators. The weak coupling expansion coefficients for $g^2$ show how well the fit approximates the data. The fit order is the order of polynomials used for the rational fit function.}
\label{tab:coefficients}
\end{table}

We also tried comparing our predictions to numerical data for the operator $S=2,J=4,n=2$ (see Fig. \ref{fig:frolov_compare} and Table \ref{tab:coefficients}). As argued before, since this operator has $n>1$, we cannot fully trust our result in this case, nevertheless the result agrees well with the numerical fits we get and the error is only slightly bigger than for the $n=1$ states. It is hard to draw conclusions about this, as there is not a lot of numerical data available for such operators.